# InSAR Phase Denoising：A Review of Current Technologies and Future Directions

Gang Xu[1], IEEE Member,　　Yandong Gao[2],　　Jinwei Li[3],　　Mengdao Xing[4], IEEE Fellow

1. State Key Laboratory of Millimeter Waves, Southeast University, Nanjing 210096, China, xugang0102@126.com.
2. School of Environment Science and Spatial Informatics, China University of Mining and Technology, Xuzhou, China.
3. Xi'an Institute of Space Radio Technology, Xi'an 710071, China.
4. National Laboratory of Radar Signal Processing, Xidian University, Xi'an 710071, China.

*Abstract*—Nowadays, interferometric synthetic aperture radar (InSAR) has been a powerful tool in remote sensing by enhancing the information acquisition. During the InSAR processing, phase denoising of interferogram is a mandatory step for topography mapping and deformation monitoring. Over the last three decades, a large number of algorithms have been developed to do efforts on this topic. In this paper, we give a comprehensive overview of InSAR phase denoising methods, classifying the established and emerging algorithms into four main categories. The first two parts refer to the categories of traditional local filters and transformed-domain filters, respectively. The third part focuses on the category of nonlocal (NL) filters, considering their outstanding performances. Latter, some advanced methods based on new concept of signal processing are also introduced to show their potentials in this field. Moreover, several popular phase denoising methods are illustrated and compared by performing the numerical experiments using both simulated and measured data. The purpose of this paper is intended to provide necessary guideline and inspiration to related researchers by promoting the architecture development of InSAR signal processing.

## I. INTRODUCTION

As a modern advanced sensor, interferometric synthetic aperture radar (InSAR) [1-7] has been successfully applied in remote sensing with great achievements and contributions. It plays an important role in topography mapping [1], deformation monitoring [3] and etc [4]. With the technology development, the interferometry has already been treated as one of the significantly important features for the current generation of remote sensors. A variety of airborne and spaceborne sensors have the interferometric capability, such as F-SAR, Shuttle Radar Topography Mission (SRTM), COSMO-SkyMed, TanDEM-X and so on. Nowadays, it is possible to provide a standard production of the global digital elevation models (DEM) with very high precision, which is also one important mission of TanDEM-X [4].

The working principle of InSAR is to measure the interferometric phase between two separate SAR images, acquired from slightly different locations. However, there is inevitable interferometric phase noise [8] introduced by some inherent factors, which can be roughly categorized as: 1) the system noise, such as thermal noise and SAR speckle noise; 2) de-correlation issues, i.e. baseline decorrelation, temporal decorrelation or volume decorrelation; 3) inaccurate signal processing, mainly involving the co-registration errors. In fact, all these factors can be viewed as de-correlation issues by introducing the phase noise. It is well known that the phase noise level is spatial-variant in the interferometric image domain, which is an inherent characteristic of InSAR system. The presence of noise increases the difficulty of phase unwrapping and even leads to phase unwrapping failure, seriously degrading the final interferometric results [9, 10]. Therefore, phase noise reduction is a necessary step during interferometric processing and it has been developed as one important technology [11-20].

From the very beginning, the originally developed method of phase noise reduction can be tracked to the multilook filter [11], which uses a strategy of simple moving average on neighbor pixels in a rectangular window, i.e. boxcar filtering [11]. Due to its easy implementation, the multilook filter has been employed during the DEM production of remote sensors, such as TanDEM-X. The obvious disadvantages of multilook filter are the resolution loss and phase fringe distortion when dealing with the high-topography and high-heterogeneity areas. Essentially, the multilook filter assumes that the interferometric phase is locally stationary and the scene reflectivity is homogeneous in a local window, where the selected samples follow the independent and identical distribution (i.i.d.). In this case, the multilook filter expects to perform a maximum likelihood (ML) estimation [12], which is also the foundation of most the phase filtering methods. However, this assumption is always not true due to the topography variation and reflectivity heterogeneity, especially when facing with the scenes of region edge, man-made structure and texture. In this case, the interferometric phase tends to exhibit the characteristics of nonstationary and nonhomogeneous, conflicting with the i.i.d. assumption. In a word, the multilook filter is just a basic non-adaptive estimator regardless of the property of interferometric phase.

Since the invention and with the development of InSAR technology, the research on phase noise reduction has been a hot topic during the last three decades. The objective of phase denoising is to accurately retrieve the interferometric phase while maintaining the spatial resolution in a possible way. Until now, there are large amounts of established and emerging algorithms with great achievements. Among these methods, the recently developed nonlocal (NL) filters have significantly attracted the attentions by exhibiting some unique superiorities [21]. Different from the traditional local filters, the NL filters measure the patch similarity to select the most relevant samples and then perform a weighted average on similar pixels for noise reduction. The sample selection strategy of patch-wise is used instead of that of pixel-wise. It is also important that the NL filters can capture the phase redundancy using the non-connected pixels, breaking the restriction of local filters. It is necessary to point out that the possible application of NL filters into TanDEM-X is studied in process /progress with some attractive results, which can be referred to [22]. Furthermore, the new concept of signal processing technologies, like sparse signal processing and machine learning, has inspired some innovate applications into InSAR phase denoising, such as sparse regularization [23], and singular value decomposition (SVD) [24]. This group methods have shown high potentials and can be treated as good candidate of a new generation in InSAR denoising processor. To boost the InSAR development, a review of phase denoising is necessary by involving the current research status. Although some literatures have roughly done a short review in their *introductions*, to the best of our knowledge, there is still no comprehensive review of phase denoising technology up till now. To fill this gap, this paper attempts to do a survey of these existing methods and update the presentation of newly proposed methods. A brief perspective on incoming and upcoming methods is addressed to exhibit their powerful potentials, pointing out the challenges and potential future directions.

## II. INTERFEROMETRIC DATA MODEL

From the view of statistical signal processing, the phase denoising is to retrieve the true value from the noisy observations, i.e. interferometric pair of SAR images. The statistical model of interferometric data and noise model of interferometric phase construct the basic foundation of InSAR technology. In order to describe the existing methods, it is necessary to introduce the signal model beforehand.

*A. Statistical model*

Let an interferometric pair of single look complex (SLC) SAR images be defined in a vector form as

$$\mathbf{k} = [z_1 \quad z_2]^T \quad (1)$$

where $z_1$ and $z_2$ are the obtained SAR images of two channels, and $T$ denotes the transpose operator. Usually, $\mathbf{k}$ can be assumed to follow a bivariate complex Gaussian distribution with zero means and its probability density function (PDF) is expressed as [25]

$$p(\mathbf{k}|\Sigma) = \frac{1}{\pi^2 \det(\Sigma)} \exp\left[-\mathbf{k}^H \Sigma^{-1} \mathbf{k}\right] \quad (2)$$

where $H$ is the transpose and conjugate operator, and $\Sigma$ is a $2 \times 2$ covariance matrix of $\mathbf{k}$, which is given by

$$\Sigma = E(\mathbf{k}\mathbf{k}^H) = \begin{bmatrix} R_1 & \rho \cdot \sqrt{R_1 \cdot R_2} \\ \rho^* \cdot \sqrt{R_1 \cdot R_2} & R_2 \end{bmatrix} \quad (3)$$

where $R_1 = E(|z_1|^2)$ and $R_2 = E(|z_2|^2)$ are the underlying reflectivities, and $\rho$ is the correlation coefficient between $z_1$ and $z_2$, which can be further decomposed as [7, 26]

$$\rho = \frac{E(z_1 \cdot z_2^*)}{\sqrt{E(|z_1|^2) \cdot E(|z_2|^2)}} = \frac{E(z_1 \cdot z_2^*)}{\sqrt{R_1 \cdot R_2}} = \gamma \cdot e^{j\phi_x}$$

$$\gamma = |\rho| = \frac{|E(z_1 \cdot z_2^*)|}{\sqrt{R_1 \cdot R_2}} \subset [0,1] \quad (4)$$

where $\gamma$ is the so-called coherence with a real value, and $\phi_x$ is the actual/noise-free interferometric phase. In (4), $I_z = z_1 \cdot z_2^*$ is defined as interferogram by the conjugate multiplication between two SAR images.

From (2) and (3), the InSAR observations, i.e. $a_1 = |z_1|$, $a_2 = |z_2|$ and $\phi_z = \angle(z_1 \cdot z_2^*)$ ($\angle(\cdot)$ is a phase operator, $\phi_z \in (-\pi, \pi]$ modulus $2\pi$), can be statistically modelled under the constraint of $R = R_1 = R_2$ and the PDF is expressed as [27]

$$p(a_1, a_2, \phi_z | R, \gamma, \phi_x) = \frac{2a_1 a_2}{\pi R^2 (1-\gamma^2)} \exp\left(-\frac{a_1^2 + a_2^2 - 2a_1 a_2 \beta}{R(1-\gamma^2)}\right) \quad (5)$$

where $\beta = \gamma \cos(\phi_z - \phi_x)$. Then, the phase distribution can be obtained by integrating $a_1$ and $a_2$ out, which is expressed as [28]

$$p(\phi_z) = \frac{1-\gamma^2}{2\pi} \cdot \left(\frac{\beta\left(\frac{1}{2}\pi + \arcsin(\beta)\right)}{(1-\beta^2)^{3/2}} + \frac{1}{1-\beta^2}\right) \quad (6).$$

The curve of $p(\phi_z - \phi_x)$ versus different values of coherence $\gamma$ is plotted in Fig.1 (a). As the increase of $\gamma$, the distribution of $\phi_z$ is much more compact with smaller noise turbulence, showing a higher quality of interferometric phase. In fact, all the noise sources aforementioned in the *introduction* can be treated as the decorrelation issues, reflecting from the decrease of coherence. As a result, lower coherence means presence of stronger noise while higher coherence indicates smaller noise level. It is well known that the phase noise level of SAR interferogram is spatial-variant, which is highly related to reflectivity and tomography of the observed scenes. Fig.2 (a) and (b) show the interferometric phase and estimated coherence using the simulated interferometric data of Mount Etna, which is used to illustrate this phenomenon.

With the InSAR development, the distribution of interferometric phase $\phi_z$ has been further completely investigated. Several researchers have independently derived the PDF of the multilook interferometric phase, which is expressed as [11]

$$p^L(\phi_z) = \frac{\Gamma\left(L+\frac{1}{2}\right)(1-\gamma^2)^L \beta}{2\sqrt{\pi}\Gamma(L)(1-\beta^2)^{L+1/2}} + \frac{(1-\gamma^2)^L}{2\pi} F\left(L, 1; \frac{1}{2}; \beta^2\right) \quad (7)$$

where $L$ is the look number, $\Gamma$ is the Gamma function, and $F$ is the Gaussian hypergeometric function. By (6) and (7), the interferometric phase distribution is unimodal, symmetric and modulus $2\pi$ with its peak at the location of actual interferometric phase $\phi_x$. By (7), the relationship of phase noise between coherence and look number is discovered and the curve of phase standard deviation with respect to coherence for specific look numbers is shown as Fig.1 (b) [11]. Moreover, the Cramer-Rao Bound (CRB) of the standard deviation of phase is also conducted as [11, 28]

$$\sigma_\phi = \frac{1}{\gamma} \frac{\sqrt{1-\gamma^2}}{\sqrt{2L}} \quad (8).$$

All these evidences indicate that increasing the coherence $\gamma$ and look number $L$ is helpful to reduce the phase noise.

For multilook filter, the increase of look number means increasing the used window size to include more available samples, assuming that all the samples in the window are stationary and homogeneous to follow i.i.d. However, this assumption is not always true for the scenes with high topography and heterogeneity. Therefore, the implementation of phase denoising needs to consider the InSAR phase characteristics.

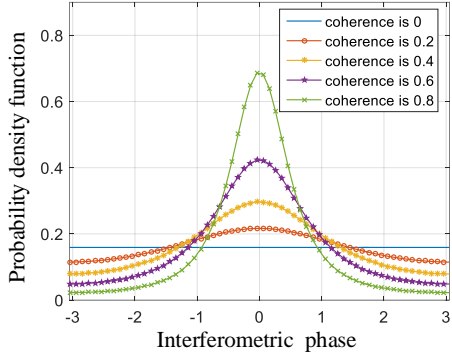

(a)

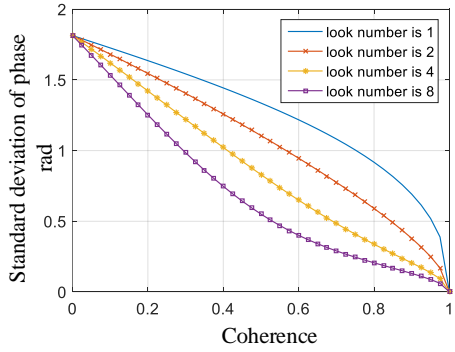

(b)

Fig.1 Statistics of interferometric phase. (a) Probability density function ($L=1$), (b) standard deviation of noisy phase.

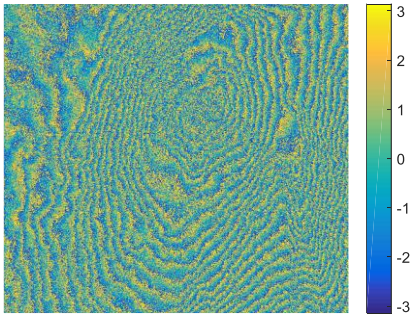

(a)

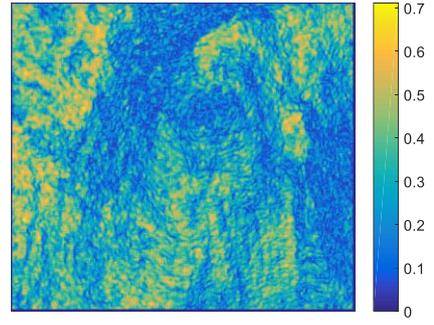

(b)

Fig.2 Simulated interferometric data of Mount Etna. (a) Noisy interferometric phase, (b) the coherence of (a).

### B. Phase noise model

In the real phase domain, the interferometric phase noise can be modelled as additive noise [13]

$$\phi_z = \phi_x + v \qquad (9)$$

where $v$ is the additive noise with zero mean. In (9), $\phi_x$ and $v$ are assumed to be independent with each other, which is an important characteristic for phase denoising. In this case, the noise standard deviation is denoted as $\sigma_v = \sigma_\phi$ and can be estimated from $\gamma$ using a lookup table shown as Fig.1 (b). As has also been aforementioned in section II-A, $v$ depends on the factors of coherence $\gamma$ and look number $L$.

Due to the phase warped property, $\phi_z$ is modulus $2\pi$ and phase unwrapping is a necessary step for the noise reduction in the real phase domain. Alternatively, phase denoising using the complex data is more effective and powerful. The noisy phase in the complex domain is modelled as [29]

$$S_z = S_x + n_v \qquad (10)$$

where $S_z = e^{j\phi_z}$, $S_x = N_c \cdot e^{j\phi_x}$ and $n_v$ are the measured phase, noise-free phase and phase noise in the complex domain, respectively. In (10), $N_c = \frac{\pi}{4} \cdot \gamma \cdot F\left(\frac{1}{2}, \frac{1}{2}; 2; \gamma^2\right)$ is determined by coherence $\gamma$, which can also be used to indicate the phase quality. The real and imaginary parts of $n_v$ are treated as zero-mean additive noise, which can be modelled as independence from the actual phase $\phi_x$. In a similar way, $n_v$ also depends on the terms of coherence $\gamma$ and look number $L$.

Last but not the least, phase denoising can also be realized from interferogram estimation/filtering, together with the de-speckling processing. As a foundation of interferogram denoising, the signal model of interferogram $I_z$ is given by [21]

$$I_z = I_x + n_I, \quad I_x = \sqrt{R_1 \cdot R_2} \cdot \gamma \cdot e^{j\phi_x} \quad (11)$$

where $I_x$ is the noise-free interferogram, and $n_I$ is the signal-dependent noise (the dependence arises from the speckle noise model, which is not focused in this paper). Under the constraint of $R = R_1 = R_2$, $I_x$ reduces to be $I_x = R \cdot \gamma \cdot e^{j\phi_x}$ while the corresponding likelihood function is denoted as (6).

Based on the aforementioned interferometric data model, a variety of InSAR phase denoising methods have been proposed in the literature. According to the history of algorithm development and technology feature, the InSAR phase denoising methods can mainly be categorized into four categories: traditional local spatial-domain filters, transformed-domain filters, NL filters and newly advanced methods. The first two categories are the groups of traditional and mature methods in this field. Fig.3 shows the statistics for the journal and conference publications of the two group methods (from web of science). It can be seen from Fig.3 that there is no significant growth in the number of publications since 2002, indicating the relatively mature research status. The NL filter can be treated as a new generation of the phase denoising technology in this field. Fig.4 shows the publication statistics of NL filters that there is significant growth since 2009 (from web of science). Moreover, as one typical advanced technique, the publication statistics of sparse methods are also shown as Fig.5 (from web of science). As a very new technology, the study on sparse technique is very limited, which needs more attention to evaluate the potentials. In the following, a non-exhaustive review is attempted to classify the existing methods, analyzing the strength and limitation of different technologies.

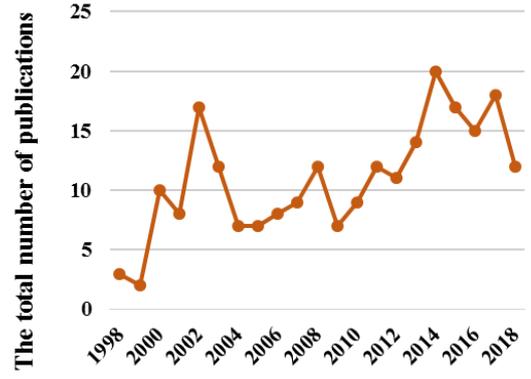

Fig.3  The statistics for total publications of local filters and transformed-domain filters.

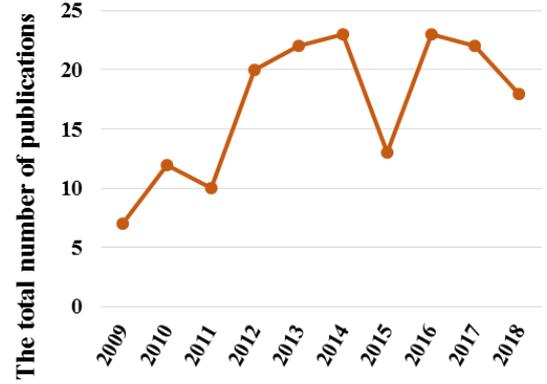

Fig.4  The statistics for publications of nonlocal filters.

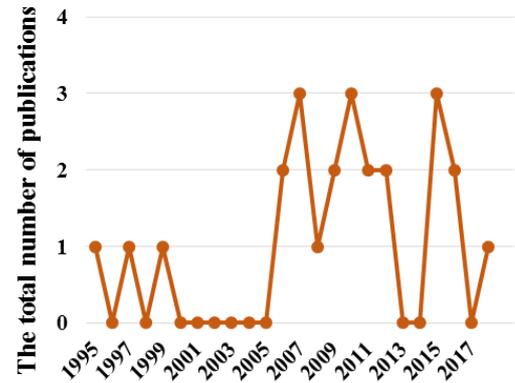

Fig.5  The statistics for publications of sparse methods.

### III.  REVIEW OF LOCAL FILTERS

As a mature research field, the local filter methods have a long history and the multilook filter [11, 30-32] can be treated as one of the very earliest methods. The basic principle of this group methods employs local window with pixels connected to select stationary and homogenous samples for phase estimation. In this section, the introduction of local filters is divided into four groups. For clarity, the processing flowchart is shown as Fig.6.

*A.  Median/mean filters*

This group methods utilize the mathematical morphologies of mean [19, 33-36] and median [16-18, 37-39] functions to estimate the phase, assuming that the local phase in a small window is statistically stationary and homogenous to follow the i.i.d. assumption. In this view, the expectation of pivoting mean/median filter is to obtain optimal filtered results from local statistics. The formulation of pivoting mean filter is expressed as

$$\tilde{\phi}_x^{mean}(k,l) = \underset{(k',l') \in win(k,l)}{mean} \left( \angle \left( \frac{S_z(k',l')}{d_{k,l}} \right) \right) + \angle(d_{k,l})$$

$$d_{k,l} = \sum_{(k',l') \in win(k,l)} S_z(k',l') \quad (12)$$

where $win(k,l)$ is a filter window centered at the site $(k,l)$, $\tilde{\phi}_x^{mean}(k,l)$ is the estimated phase of pixel $(k,l)$, and $mean(\cdot)$ is an average operator. Meanwhile, the formulation of median filter is given by

$$\tilde{\phi}_x^{median}(k,l) = \underset{(k',l') \in win(k,l)}{median} \left( \angle \left( \frac{S_z(k',l')}{d_{k,l}} \right) \right) + \angle(d_{k,l}) \quad (13)$$

where $median(\cdot)$ is a median operator. The preconditions of both the filters highly depends on the stationary and homogenous property of local phase that the topography fluctuation in the local region is slow for the sampling rate and the adjacent pixels have high correlation. When the noise is additive Gaussian noise, the pivoting mean filter is statistically optimal from a view of ML estimation. However, the stationary and homogenous assumption conflicts with the fast variation of topography. Similar to the multilook filter, the mean filter unfortunately has over-smoothing effect and can hardly deal well with the phase details of high topography variation. Compared with the mean filter, the pivoting median form has a better preservation of phase fringe, but the power of noise suppression reduces in some degree. Besides, one major limitation of this group methods is non-adaptive regardless of the local spatial-variation of the selected samples, such as noise level, using an equal average. Several modifications have been designed with improvement by distinguishing the contributions between different samples, such as weighted average [36, 39].

### B. Lee filter and its refined forms

The basic principle of this group methods uses locally directional window and applies Lee filter [13] or refined version [14, 40, 41] with adaptive local noise adjustment. The processing of Lee filter can be divided into two steps. The first step is the stationary and homogenous sample selection by evaluating from candidates of different directional windows. To adaptively capture the phase fringe, Lee filter has predesigned different windows of 16 directions and tries to choose the best association one. The next step is statistical parameter estimation and minimum mean square error (MMSE) estimation. The generalized mathematical formulation of Lee filter is shown as below

$$\tilde{\phi}_x = \bar{\phi}_z + \eta \cdot (\phi_z - \bar{\phi}_z)$$

$$\eta = \frac{var(\phi_x)}{var(\phi_z)} = \frac{var(\phi_z) - \sigma_v^2}{var(\phi_z)} \quad (14)$$

where $\tilde{\phi}_x$ is the estimation of $\phi_x$, $\bar{\phi}_z$ is the mean value of $\phi_z$ in the filtering window, $\eta \subset [0,1]$ is a weighted coefficient of Lee filter and as aforementioned that $\sigma_v$ can be estimated from $\gamma$ using a lookup table. To calculate $\bar{\phi}_z$ and $var(\phi_z)$, $\phi_z$ needs to be unwrapped in the filtering window. As an alternative way, the complex formulation of Lee filter is given by

$$\tilde{S}_x = \bar{S}_z + \eta \cdot (S_z - \bar{S}_z) \quad (15)$$

where $\tilde{S}_x = e^{j\tilde{\phi}_x}$, $S_z = e^{j\phi_z}$ and $\bar{S}_z$ is the normalized mean value of $S_z$ in the filtering window. In the area of high coherence, the value of $\eta$ is very close to 1 and Lee filter simplifies to be $\tilde{\phi}_x \approx \phi_z$, avoiding the smoothing effect. In the area of low coherence, Lee filter reduces to be a mean filter as $\tilde{\phi}_x \approx \bar{\phi}_z$ when $\eta = 0$. The performance of Lee filter highly depends on the accuracy of directional window selection and parameter estimation. The major drawback of Lee filter is the number limitation of predefined window and sensitivity to strong noise, which has limitation when facing with the phase fringe of complicated structures and textures. In the area of high topography, Lee filter does not work well and tends to introduce some undesired artifacts, such as discontinuity. Several modifications have been designed to Lee filter with improvements, such as local adaptive filter [15, 40] and refined Lee filter [14]. To overcome the number limitation of window, the strategy of adaptively estimating the directional window is employed by retrieving the local frequency to determine the direction of phase fringe [15]. The refiled Lee filter proposed in [14] has done some modifications to improve the robustness of directional window selection and statistical parameter estimation.

*C. Local frequency estimators*

The local frequency estimators are based on the polynomial phase model, such as linear [42-45] and nonlinear [46-49]. The basic principle of this group methods estimates the local frequency of the interferometric phase to derive the phase fringe. Usually, the local phase in a small window is modelled as sine wave with one major component, which is shown as

$$S_z^{win}(k,l) \approx a_{k,l} \cdot \exp\left[j2\pi(k \cdot f_x + l \cdot f_y)\right] + n_v(k,l) \quad (16)$$

where $S_z^{win}(k,l)$ is a local window form of $S_z(k,l)$ centered at site $(k,l)$, $a_{k,l}$ and $(f_x, f_y)$ are the coefficient and two-dimensional (2-D) frequency, respectively. Then, $(f_x, f_y)$ can be estimated using the spectrum estimators, such as ML [42], vector covariance matrix (VCM) [42], and multiple-signal classification (MUSIC) [45]. Based on the estimated fringe frequencies, the filtered phase can be estimated as

$$\tilde{\phi}_x(k,l) = \angle\left(\sum_k\sum_l S_z^{win}(k,l) \cdot \exp\left[-j2\pi(k \cdot \tilde{f}_x + l \cdot \tilde{f}_y)\right]\right) \quad (17)$$

where $(\tilde{f}_x, \tilde{f}_y)$ is the estimated 2-D frequency. In the practice, the linear frequency model is not accurate enough by neglecting other frequency components. To overcome this limitation, several innovative strategies, such as iteration with varying the window size [18] and nonlinear phase modelling [47-49], are applied to preserve the phase details of fringe and texture. The order of model selection is crucial to this group methods. The higher the order choose, the more accurate the phase estimation. However, the cost is the increase of computational load. As a result, it needs to balance between order selection and computational complexity. Besides, the integration methods of phase frequency estimation and unwrapping have also been studied [51-57], which is not introduced here.

*D. Hybrid methods*

The procedure of this group methods is usually decomposed into two step estimations. The basic idea is that the interferometric phase can be separated into the principal and residual components, such as low-resolution (LR) fringe and high-resolution (HR) pattern [18, 41, 60, 61]. The principal component can usually be extracted using the technique of local frequency estimation and so on. After principal phase compensation, the residual component of interferometric phase can be estimated more precisely. It is obvious that the phase is estimated in a hybrid way, which is shown as

$$\tilde{\phi}_x = \tilde{\phi}_x^{LR} + \tilde{\phi}_x^{HR} \quad (18)$$

where $\tilde{\phi}_x^{LR}$ and $\tilde{\phi}_x^{HR}$ are the estimations of the principal and residual phase, respectively. For the local filters, the sample selection strategy, such as the window form, is critical to the phase/frequency estimation. Until now, the developed window can be roughly classified as oriented/directional window [13-15], adaptive size-varying window [18, 43] and region growing mask [59, 60]. It is highlighted that the region growing (named as intensity-driven adaptive neighborhood, IDAN) method [60] employs local window with adaptive shape and size to search for the similar pixels of being stationary and homogenous. Among the traditional local filters, the region growing method can select the most relevant samples with high quality, which can be treated as one best of the established sample selectors.

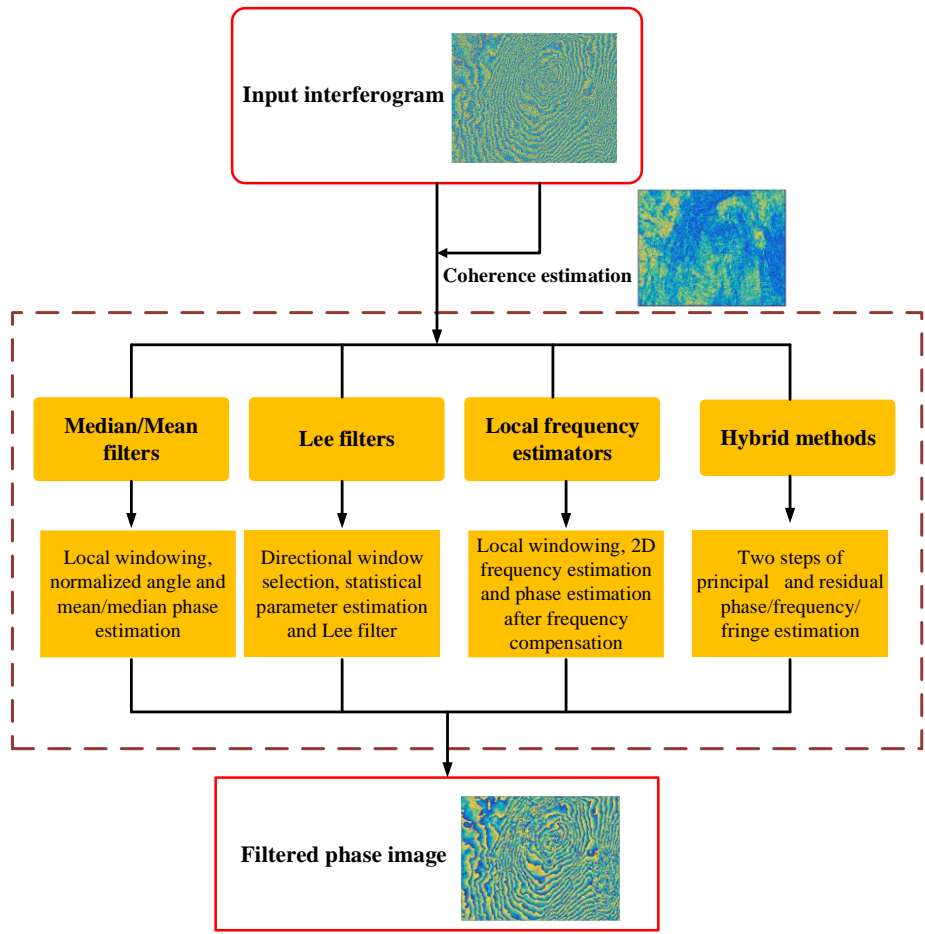

Fig.6 The processing flowchart of the local filters.

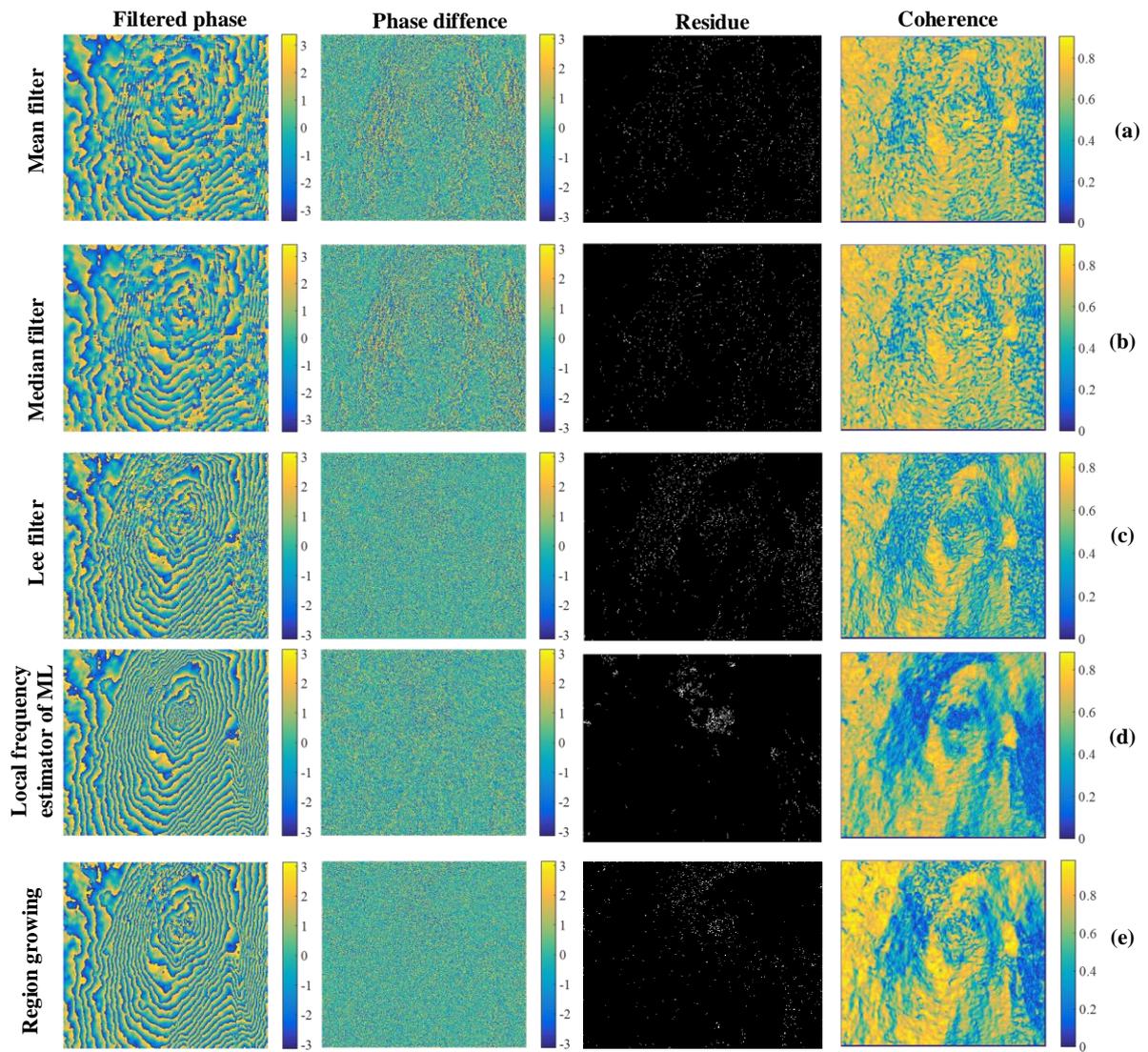

Fig.7 Filtered phase images using local filters. (a) Pivoting mean filter [19] (residue number 2831), (b) pivoting median filter [39] (residue number 2826), (c) Lee filter [13] (residue number 6161), (d) local frequency estimator using ML method [42] (residue number 2409), (e) region growing method [60] (residue number 2189).

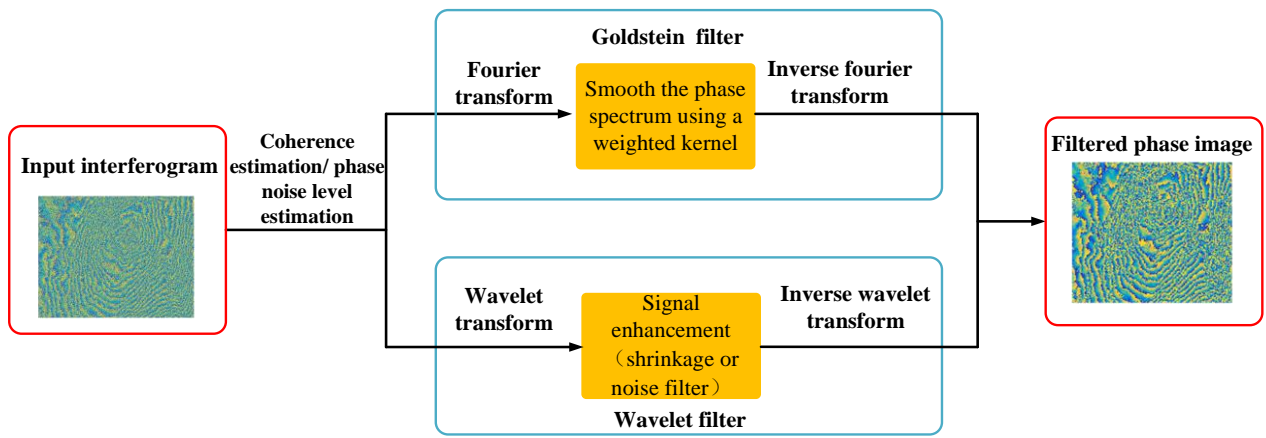

Fig.8 The processing flowchart of the transformed-domain filters.

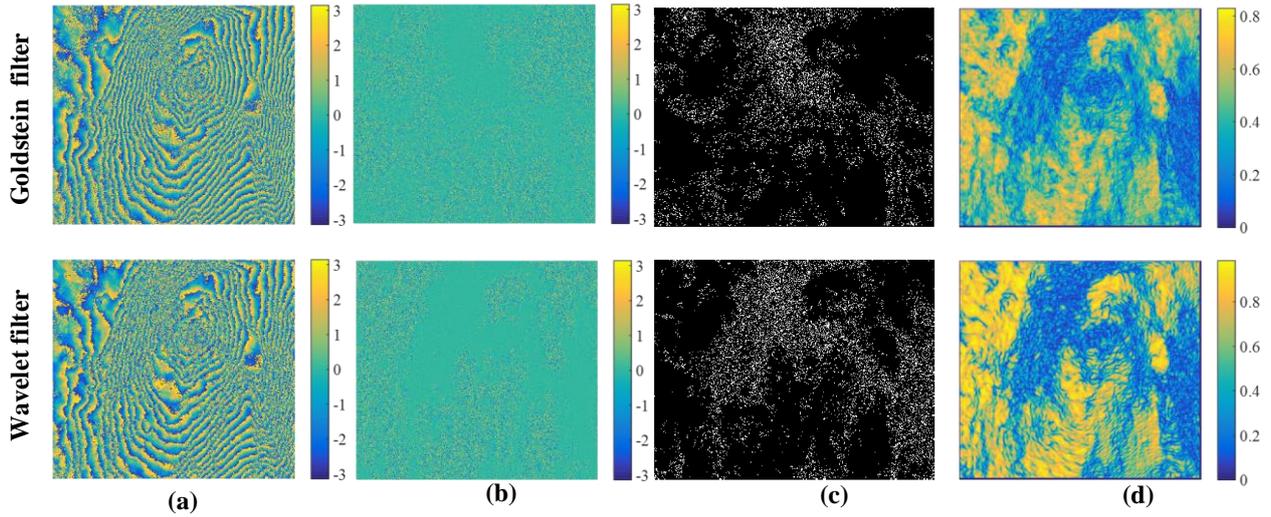

Fig.9 Filtered phase images using the transformed-domain filters. The first and second rows are Goldstein filter [62] and wavelet filter [29], respectively.

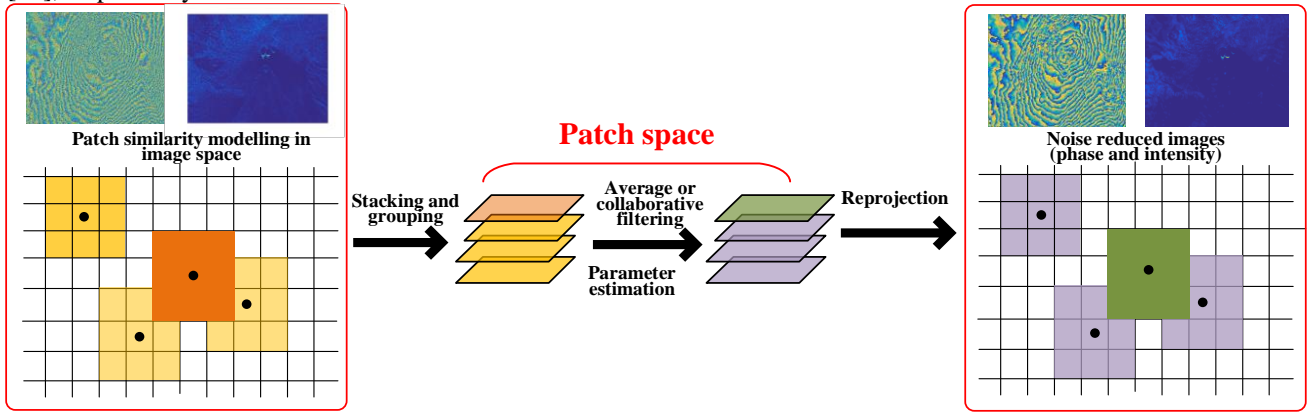

Fig.10 The processing flowchart of the nonlocal filters.

### E. Comparison of local filters

In this subsection, numerical experiments using the aforementioned interferometric data are performed to compare some typical local filters, i.e. pivoting mean filter [19], pivoting median filter [39], Lee filter [13], ML method [42], and region growing method [60]. Fig.7 shows the filtered images that the rows are the filtered phase, phase difference (between filtered and noisy ones), the residue and coherence, respectively. It can be seen that the pivoting mean and median filters tend to introduce phase discontinuity in the highly-sloped area, producing many residues. The ML method has a better performance in the high topography regions. However, this method will produce excessive filtering, which will lead to loss of phase details. Besides, the size of the filter window highly affects the ML results. In comparison with the three aforementioned filters, Lee filter has a better performance on noise reduction and detail preservation. However, Lee filter still does not work well by introducing artifacts in the area of very low coherence. Among these filters, the performance of region growing method is nearly the best. However, the improvement of phase estimation in the highly-sloped regions is still very limited. It can be concluded that the traditional local filters are not very effective to deal with the high topography variation, which is an inherent limitation of the local filters.

## IV. REVIEW OF TRANSFORMED-DOMAIN FILTERS

Instead of filtering in the spatial domain, the InSAR phase denoising can also be performed in the transformed domain. The transformed-domain filters have some superiorities because the phase signatures can be enhanced to be more separable from the noise in the transformed domain. In this section, we introduce the transformed-domain filters and divide them into two groups. For clarity, the processing flowchart of the transformed-domain filters is shown as Fig. 8.

### A. Frequency domain methods

The frequency domain usually refers to the Fourier transform and the discrete cosine transform. The most typical and effective one is the Goldstein filter [62] and its refined versions [63-74]. The basic idea lies into the fact that the interferometric phase is locally stationary and homogeneous and most of the useful components of local phase is usually limited to a narrow band while the noise is distributed in the entire frequency band. The Goldstein filter transfers the interferometric phase image into frequency domain and weights the phase spectrum for noise reduction. The generalized formulation of Goldstein filter is expressed as

$$\tilde{S}_x(k,l) = \text{IFFT2}\left(\text{SM}\left(\left|Y_z^{win}(p,q)\right|\right)^{\alpha} \cdot Y_z^{win}(p,q)\right) \quad (19)$$

where $Y_z^{win}$ is the Fourier spectrum of the interferometric phase image in a local window centered at site $(k,l)$, $\text{IFFT2}(\cdot)$ is the 2-D inverse Fourier transform, $\text{SM}(\cdot)$ represents the smoothing operation, and $\alpha$ is the filtered parameter. In fact, proper selection of $\alpha$ is very important for the Goldstein filter. The performance of Goldstein filter is greatly affected by the patch size of local window and filter parameter $\alpha$. When $\alpha = 0$, it means no filter power without any processing. When $\alpha = 1$, the Goldstein filter has the strongest filtering power. To improve the robustness, Baran [63] has defined the filtering parameter as $\alpha = 1-\gamma$, meaning that the coherence reflects the noise level to adaptively control the filtering strength. Furthermore, several improvements have been achieved by providing more moderate estimation of the filtering parameter [67], adaptive window size [65], topography compensation [64] and etc [70-74].

### B. Wavelet domain filters

The wavelet transform can enhance the signatures of interferometric phase using the multiresolution representation. The multiresolution property has the advantages of reliable noise separation and perfect nonstationary signal analysis. Lopez and Fabregas propose to apply wavelet transform for phase denoising by modelling the phase noise in the wavelet domain [29], which provides a foundation of this group methods. Based on the assumption of linear wavelet transform, the additive phase noise model in (10) can be easily generated to the wavelet domain. The wavelet transform of (10) can be expressed as

$$W_z = W_x + W_v \quad (20)$$

where $W_z$, $W_x$ and $W_v$ are the wavelet transforms of $S_z$, $S_x$ and $n_v$, respectively. In the $i$ th scale of 2-D wavelet transform, the detailed form of $W_x$ is given by

$$W_x(k,l,i) = N_c \cdot 2^i e^{j\phi_x^w(k,l,i)} \quad (21)$$

where $\phi_x^w(k,l,i)$ represents the phase information at site $(k,l)$ of $i$ th wavelet scale. In the term of noise, the variance is not change before and after the wavelet transform. When one deep scale of wavelet transform is applied, the wavelet transform multiplies the phase term by a factor of two without altering the noise level [29]. The behavior of wavelet transform can effectively locate, preserve and amplify the actual phase in the wavelet domain. In this way, the phase and noise can be more easily separated. Then, the phase noise reduction is performed by means of appropriate nonlinear processing, such as wavelet shrinkage or filtering, which can be expressed in a generalized formulation as

$$\tilde{W}_x = H(W_z) \cdot W_z \quad (22)$$

where $H(\cdot)$ denotes a shrinkage or filter function. After all, the inverse wavelet transform is applied to $\tilde{W}_x$ and the phase image can be obtained with noise reduction. The performance of wavelet filter depends on the scale of wavelet decomposition and the shrinkage/filter function used in the wavelet domain. For better multiresolution capability, the undecimated wavelet transform [76] and wavelet packets [77, 78] are used to improve the phase denoising performance. It is commonly accepted that the wavelet-transformed filters have good preservation of spatial resolution due to the perfect representation of nonstationary signals from multiresolution property.

## C. Comparison of transformed-domain filters

In this subsection, the experiments using the same dataset are carried out to show the performances of transformed-domain filers. Fig.9 shows the results using the Goldstein and wavelet filters (residue numbers are 18620 and 25289), respectively. For the Goldstein filter [62], the parameter is set as $\alpha = 0.999$ with a strong filtering effect. As Fig.9 shows that the Goldstein filter has a good phase denoising performance in the high coherence area. However, there are so many residues and large phase differences in the low coherence area, indicating the phase distortion in the high topography areas. For the wavelet filter [29], the dual-tree complex wavelet transform is used for excellent phase preservation. It can be seen from Fig.9 that, in comparison with the Goldstein filter, the wavelet filter has a better performance of phase detail preservation because the wavelet coefficients of noise are still remained with only amplifying the signal components.

## V. REVIEW OF NONLOCAL FILTERS

As a new generation, the technology of NL filter has emerged high success in InSAR phase denoising, which has been a hot research topic now. The paradigm of NL filter lies into the sample selection strategy of measuring the patch similarity without restricting the search window in a local region, overcoming the limitation of neighbor connection. In the image processing, the concept of NL filter can be tracked to the pioneering work of Lee [79] and the following seminal work of Buades [80, 81] has developed the patch-based filter framework. With the technology development, the NL filter has been attempted into InSAR by establishing new technology, known as NL-InSAR [21], which can be treated as one of the most typical and effective methods. The NL filters employ the patch-wise method to measure the patch similarity in the interferogram domain and weighted average is applied to estimate the filtered interferogram together with phase denoising. In this framework, the most relevant samples can be selected with high quality and phase denoising is achieved by distinguishing the contribution between different samples. It is commonly accepted that the NL filters have the advantage of phase fringe/texture preservation by effectively capturing the phase structure and extracting the phase redundancy in the patch space. In comparison with the traditional local filters, the NL filters are more power with some unique superiorities, such as better performance in dealing with the high topography and high heterogeneity (such as isolated scenes). For clarity, the processing flowchart of NL filters is shown as Fig. 10.

*Concept of nonlocal filters:* the seminal work of Buades [80, 81] has established the framework of NL filter by means of patch-based method and weighted average strategy. To estimate the noise free interferogram at site $u$, the generalized formulation of NL filter is expressed as

$$\tilde{I}_x^u = \frac{\sum_t \omega_{u,t} \cdot I_z^t}{\sum_t \omega_{u,t}}, \omega_u = \sum_t \omega_{u,t} \quad (23)$$

where $\omega_{u,t}$ is the weighted coefficient by considering the contribution from the pixel at site $t$. The key idea of NL filter is that the weight $\omega_{u,t}$ is measured from the similarity between the patches $\mathbb{C}_u$ (centered at site $u$) and $\mathbb{C}_t$ (centered at site $t$), known as patch-based method. Usually, $\omega_{u,t}$ is designed using an exponential kernel of the patch similarity parameter $\Delta$ as

$$\omega_{u,t} = \exp\left[\frac{\Delta}{h}\right] \quad (24)$$

where a large value of $\Delta$ means stronger patch similarity to place large weight $\omega_{u,t}$, and $h$ controls the decay of the exponential function. Essentially, the choice of $h$ is a tradeoff between filtering strength and detail preservation. From (24), a choice of large value $h$ leads to less discriminative weight with stronger filtering strength while a smaller one enhances the patch dissimilarity for better detail preservation.

The design of the weights $\omega_{u,t}$ is an important issue of the NL filter, which is associated to the patch similarity. So it is necessary to properly model the similarity criteria for the NL filters of InSAR, which is crucial to the phase denoising performance. Besides, the transform-domain filter is also helpful for phase denoising. It is very interesting to combine the NL principle in combination with the transform-domain filter, such as wavelet transform. Accordingly, the review of NL filters is divided into two main groups in this section.

### A. Probabilistic patch-based filters

The probabilistic patch-based (PPB) filter [21] extends the NL filter to the application of SAR/InSAR using the weighted maximum likelihood estimator (WMLE) [81]. Based on the statistical model of interferometric data in (5), the generalized formulation of PPB filter for interferogram denoising is shown as [21]

$$\tilde{\Theta}_u = \arg\max_{\Theta_u} \sum_t \omega_{u,t} \cdot \log p(O_t|\Theta_u) \quad (25)$$

where $\Theta_u = (R_u, \gamma_u, \phi_x^u)$ is a set of desired parameters for pixel at site $u$ and $O_t = (a_1^t, a_2^t, \phi_z^t)$ is the observations of pixel at site $t$. The detailed form of InSAR parameters estimation is given by

$$\tilde{\phi}_x^u = \angle \tilde{I}_x^u, \quad \tilde{\gamma}_u = \frac{|\tilde{I}_x^u|}{\tilde{R}_u}$$
$$\tilde{R}_u = \sum_t \frac{\omega_{u,t}}{\omega_u} \cdot \frac{|z_1^t|^2 + |z_2^t|^2}{2} \quad (26).$$

For the PPB filter in [21], the patch similarity is determined as: the patch similarity can be treated as measuring how likely the two patches $\mathbb{C}_u$ and $\mathbb{C}_t$ follow the same distribution. Based on the independence assumption of the pixels in each patch, the patch similarity can be computed pixel-wise. The weight $\omega_{u,t}$ is defined as [21]

$$\omega_{u,t} \overset{\Delta}{=} p(O_{\mathbb{C}_u}, O_{\mathbb{C}_t} | \Theta_{\mathbb{C}_u} = \Theta_{\mathbb{C}_t})^{1/h}$$
$$= \prod_m p(O_{\mathbb{C}_u(m)}, O_{\mathbb{C}_t(m)} | \Theta_{\mathbb{C}_u(m)} = \Theta_{\mathbb{C}_t(m)})^{1/h} \quad (27)$$

where $\mathbb{C}_u(m)/\mathbb{C}_t(m)$ is the $m$th pixel of the patch $\mathbb{C}_u/\mathbb{C}_t$. Accordingly, it also means that the similarity parameter $\Delta$ is computed as [21]

$$\Delta_{u,t}^1 = \sum_m \log p(O_{\mathbb{C}_u(m)}, O_{\mathbb{C}_t(m)} | \Theta_{\mathbb{C}_u(m)} = \Theta_{\mathbb{C}_t(m)}) \quad (28)$$

It is also important that the PPB filter employs a strategy of iterative manner. Using the intermediate results, the estimation accuracy of weight computation can be enhanced by using a posterior probability as

$$p(\Theta_{\mathbb{C}_u} = \Theta_{\mathbb{C}_t} | O)$$
$$\propto p(O_{\mathbb{C}_u}, O_{\mathbb{C}_t} | \Theta_{\mathbb{C}_u} = \Theta_{\mathbb{C}_t}) \cdot p(\Theta_{\mathbb{C}_u} = \Theta_{\mathbb{C}_t}) \quad (29)$$

where $p(\Theta_{\mathbb{C}_u} = \Theta_{\mathbb{C}_t})$ is a prior term. Then, the similarity parameter $\Delta$ is refined as [21]

$$\Delta_{u,t}^2 = \sum_m \log p(\Theta_{\mathbb{C}_u(m)} = \Theta_{\mathbb{C}_t(m)} | O) \quad (30).$$

In this way, the weights can be refined at each iteration. The performance of PPB filter relies on the accuracy of parameter estimation, i.e. $\Theta_{\mathbb{C}_u(m)}$ and $\Theta_{\mathbb{C}_t(m)}$. The presence of phase topography and scene heterogeneity conflicts with the collection of relevant samples. In this case, the PPB filter may produce over-smoothing in the areas of high topography and high heterogeneity. Several modifications [22, 82-84], such as topography compensation and heterogeneity measurement, have been designed to improve the reliability of the PPB filter in interferogram denoising. Besides, due to the iterative manner of PPB approach, the fast solution of NL filter has also been studied on the GPU platform to consider the practical application [85].

### B. Block matching 3-D filters

The InSAR block matching 3-D (BM3D) filter [86] modifies the traditional BM3D filter into the application of InSAR domain. The concept of InSAR BM3D filter can be decomposed into three main steps: 1) use patch-based method to collect the similar patches, and group/stack these similar patches; 2) wavelet transform on the 3-D block data, phase filtering in the wavelet domain and then inverse wavelet transform back to the patch domain; 3) return all the filtered patches to the original locations and aggregate them with overlapped data fusion. The basic principle of noise reduction lies into the fact that the 3-D wavelet transform can enhance the signatures of interferometric phase, providing more reliable noise separation. The formulation of 3-D wavelet transform on patch block data is expressed as

$$W_{I_z} = W_{I_x} + W_{n_I} \quad (31).$$

The phase noise reduction in the wavelet domain is decomposed into two steps, where the first step is the hard thresholding formulated as

$$\tilde{W}_{I_x}^u = \begin{cases} W_{I_z}^u & |W_{I_z}^u| > T \\ 0 & \text{otherwise} \end{cases} \quad (32)$$

where $W_{I_z}^u$ denotes the pixel of $W_{I_z}$ at site $u$, $T$ is the threshold, and the estimation of $W_{I_x}^u$ is refined in the second step using Wiener filtering as

$$\max\left(\frac{\left(\tilde{W}_{I_x}^u\right)^2 - \sigma_W^2}{\left(\tilde{W}_{I_x}^u\right)^2}, 0\right) \cdot \tilde{W}_{I_x}^u \qquad (33)$$

where $\sigma_W^2$ is the noise variance in the wavelet domain. Therefore, the InSAR BM3D filters have both the superiorities of NL filter and wavelet-domain filter. In addition to the wavelet-domain filter, some other advanced methods have also been proposed, such as pyramidal representation [87], higher order SVD [24] and etc [88-92].

*Experimental analysis:* As a typical method, the NL-InSAR algorithm proposed in [21] is applied to illustrate the NL filter performance. First, the experiments are performed using the aforementioned data and the filtered images are shown as Fig.11. The search window seize is $15 \times 15$, the patch size is $5 \times 5$ and $h = 12$. As Fig. 11 shows that the NL-InSAR filter deals well with the smooth area with little disturbance, however, there is artifacts for the filtered phase in the high heterogeneity and high-slopped areas. Then, more experiments are implemented using the RADARSAT-2 repeat-pass interferometric sample data. Due to the resolution limitation, the search window seize is $11 \times 11$, the patch size is $3 \times 3$ and $h = 6$. Fig. 12 shows the filtered images that the NL-InSAR filter is very powerful of phase and amplitude denoising in the application of building areas.

## VI. REVIEW OF NEWLY ADVANCED METHODS

With the recent development of new concept of signal processing, such as sparse signal processing and machine/deep learning [93, 94], some advanced methods have been proposed to do attempt in InSAR phase denoising. Among these methods, the group of sparse methods [95, 96] has drawn a lot of attentions to show potentials in some degree. In a high probability, this group methods may be treated as good candidate of next technology generation of InSAR phase denoising. In this section, a review of the sparse methods is introduced.

The sparse methods can retrieve the noise-free image by applying a sparse constraint in a generalized formulation as

$$\min_{\mathbf{x}/\mathbf{s}, \Psi} \|\mathbf{s}\|_0, \quad \text{subject to} \begin{cases} \mathbf{x} = M(\mathbf{z}_1, \mathbf{z}_2) \\ \mathbf{s} = \Psi(\mathbf{x}) \end{cases} \qquad (34)$$

where $\|\cdot\|_0$ is the $L_0$-norm of a vector, $\mathbf{x}/\mathbf{s}$ is the noise-free image to be recovered, containing the actual phase $\phi_\mathbf{x}$, $\Psi$ is a dictionary or operator to sparsely represent $\mathbf{x}$, $\mathbf{z}_1$ and $\mathbf{z}_2$ are the vectors of interferometric pair of SAR images $z_1$ and $z_2$, respectively. In (34), $\mathbf{x} = M(\mathbf{z}_1, \mathbf{z}_2)$ is a linear or nonlinear function by mapping $\mathbf{z}_1$ and $\mathbf{z}_2$ to $\mathbf{x}$, which can be treated as a uniform formulation of (9), (10) and (11). Accordingly, the sparse technology can perform noise reduction on real phase, complex phase and interferogram, respectively. In fact, the direction solution of (34) is very difficult because it is a NP-hard problem. Meanwhile, it also depends on the forms of $M$ and $\Psi$ when solving (34). In the following, two typical groups of sparse methods are discussed, considering their contributions.

### A. Sparse regularization

This group of sparse methods employs the Bayesian rules to restore the filtered image through a maximum a posteriori (MAP) estimator [97-100]. Usually, the MAP estimation is transferred to solve an energy minimum problem, which can be expressed in a generalized formulation as

$$\begin{aligned} &\min_{\mathbf{x}} E(\mathbf{z}_1, \mathbf{z}_2, \mathbf{x}) \\ &E(\mathbf{z}_1, \mathbf{z}_2, \mathbf{x}) = \Phi(\mathbf{z}_1, \mathbf{z}_2, \mathbf{x}) + \kappa \cdot \|\Psi(\mathbf{x})\|_1 \end{aligned} \qquad (35)$$

where $\Phi(\mathbf{z}_1, \mathbf{z}_2, \mathbf{x})$ is the data fidelity term from the likelihood function, $\|\Psi(\mathbf{x})\|_1$ is the $L_1$-norm term using a sparse prior, and $\kappa$ is a regularized coefficient to balance the two terms. In (35), $\Phi(\mathbf{z}_1, \mathbf{z}_2, \mathbf{x})$ can usually have three different formulations according to the noise reduction in different domains. Take the interferogram regularization as an example, $\Phi(\mathbf{z}_1, \mathbf{z}_2, \mathbf{x})$ can be derived from a joint likelihood of (5) as

$$\Phi(\mathbf{z}_1, \mathbf{z}_2, \mathbf{x}) = \sum_{a_1 \in |\mathbf{z}_1|, a_2 \in |\mathbf{z}_2|, \phi_z \in \phi_\mathbf{z}} 2\log R + \frac{a_1^2 + a_2^2 - 2a_1 a_2 \beta}{R(1 - \gamma^2)} \qquad (36)$$

where $\mathbf{x}$ is composed of $R$ and $\phi_x$ (involved in $\beta$ shown as (4) and (5)). On the other hand, the format of $\|\Psi(\mathbf{x})\|_1$ highly determines the regularization power. There are many different sparse regularization approaches, such as total variation (TV) and wavelet-domain regularizations, which are listed as follows

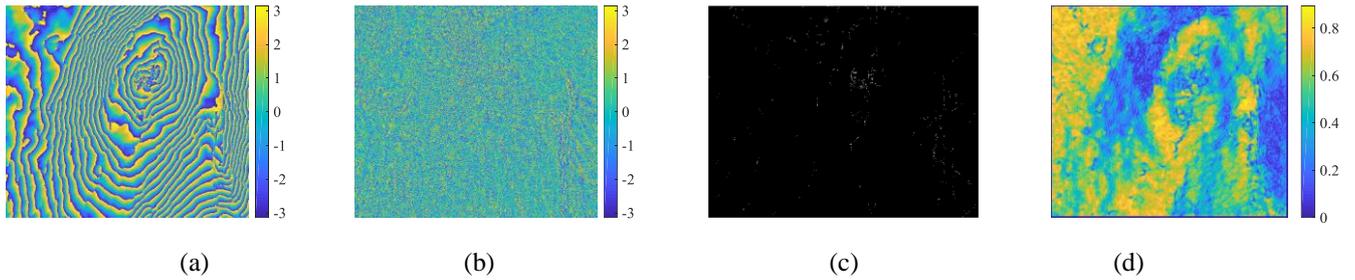

Fig.11 Filtered phase images using the NL-InSAR filter [21] (Mount Etna data). (a) Filtered phase, (b) phase difference, (c) residues, (d) coherence of filtered images.

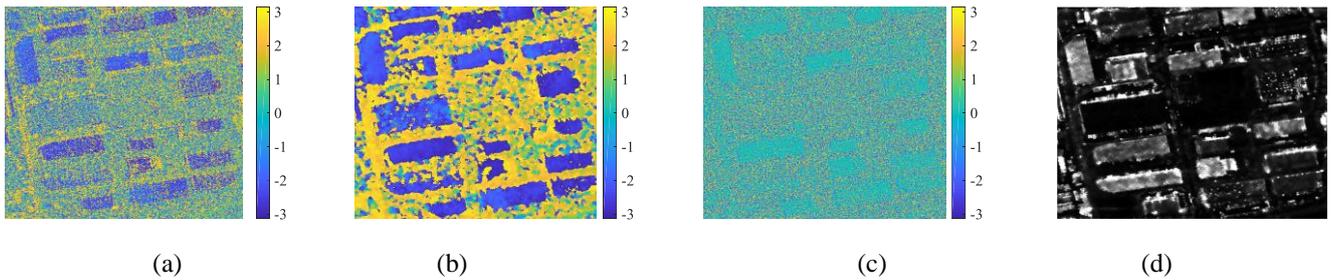

Fig.12 Filtered images using the NL-InSAR filter [21] (RADARSAT-2 data). (a) Original phase, (b) filtered phase, (b) phase difference, (d) filtered amplitude.

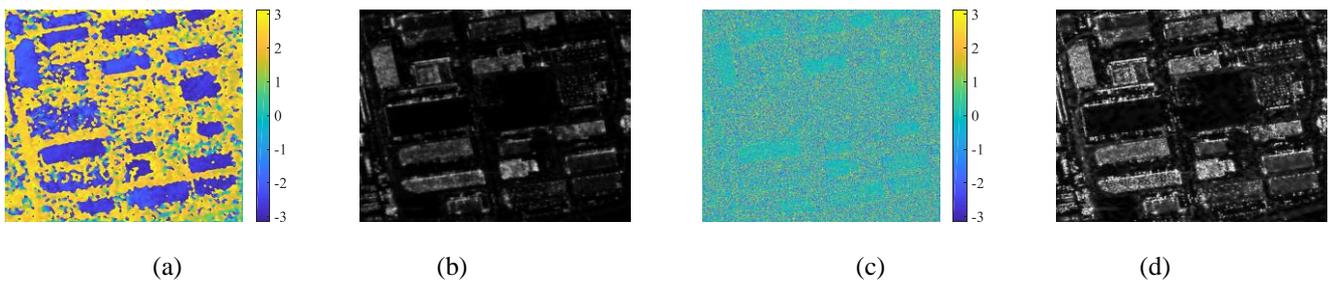

Fig.13 Filtered images using the sparse methods (RADARSAT-2 data). TV regularization [23]: (a) filtered phase, (b) filtered amplitude. Wavelet regularization [100]: (c) filtered phase, (d) filtered amplitude.

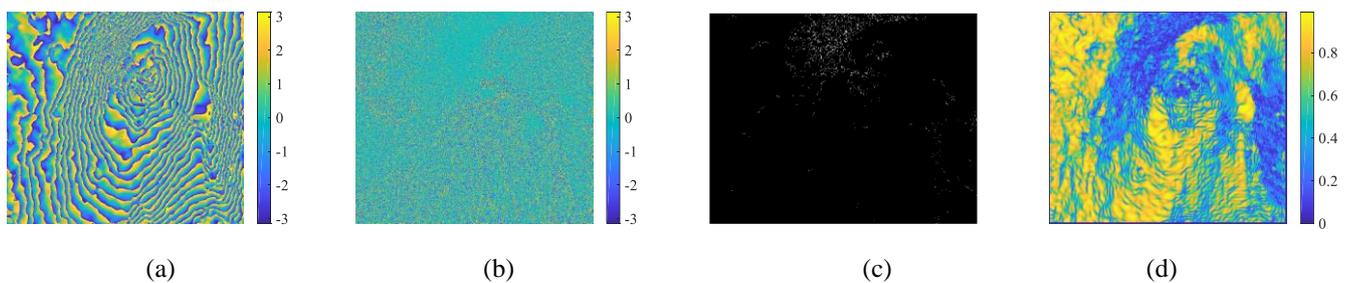

Fig.14 Filtered phase images using the sparse wavelet regularization [100].(a) Filtered phase, (b) phase difference, (c) residues (3773), (d) coherence of filtered images.

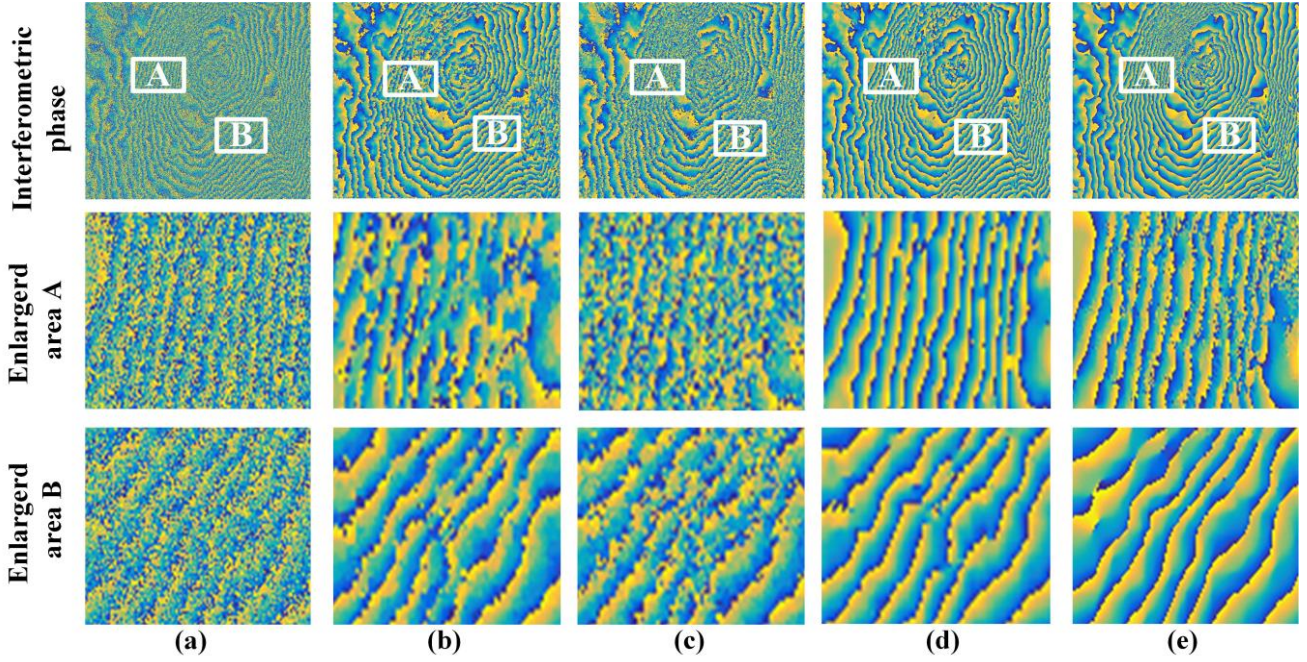

Fig.15 Filter phase images using methods of different categories (Mount Etna data) [100]. (a) Original, (b) Lee filter [13]. (c) wavelet filter [29], (d) region growing [60], (e) sparse wavelet regularization [100].

*1) TV regularization* [23, 98]. Usually, the TV regularization applies the gradient operation on the 2-D phase image for noise reduction, which is expressed as

$$\|\Psi(\mathbf{x})\|_1 = \sum_u \left|\nabla\left(\phi_x^u\right)\right| \tag{37}$$

where $\nabla\left(\phi_x^u\right)$ denotes the 2-D gradient operator on phase $\phi_x$ at site $u$. The phase denoising is beneficial from the local smooth effect using gradient operation. To improve the robustness of phase reduction, joint TV regularization of phase and amplitude of InSAR data has also been studied, where the regularization is formulated as

$$\|\Psi(\mathbf{x})\|_1 = \sum_u \max\left(\left|\nabla\left(\sqrt{R^u}\right)\right|, \lambda \cdot \left|\nabla\left(\phi_x^u\right)\right|\right) \tag{38}$$

where $\sqrt{R^u}$ denotes the pixel of amplitude image $R$ at size $u$, and $\lambda$ is used to balance the regularization between amplitude and phase. In fact, the sparse TV regularization is not very suited to phase denoising when dealing with the scenes of high topography and high heterogeneity, not satisfying the sparsity assumption on phase gradient.

*2) Wavelet-domain regularization* [100, 101]. The generalized formulation of sparse wavelet regularization is given by

$$\|\Psi(\mathbf{x})\|_1 = \sum_u \left|W_x^u\right| \tag{39}$$

where $W_x^u$ denotes the pixel of wavelet image $W_x$ at site $u$. In this framework, the phase noise reduction is achieved by sparse regularization of phase in the wavelet domain. Besides, the amplitude and phase can be jointly regularized in the wavelet domain to improve the sparsity degree, which is helpful to noise reduction. Accordingly, both the amplitude and phase noise reduction can be achieved using this approach.

### B. Sparse coding

Different from sparse regularization using a fixed dictionary, the group methods of sparse coding involve the dictionary learning of $\Psi$ to improve the sparse representation of phase image. The generalized formulation is shown as [102-104]

$$\min_{\mathbf{s},\Psi} \sum_u \|\mathbf{y}_u - \Psi \mathbf{s}_u\|_2^2 + \kappa_u \cdot \|\mathbf{s}_u\|_1 \tag{40}$$

where $\mathbf{y}_u$ is a vector of one patch of InSAR observations (i.e. $S_z/I_z$) centered at site $u$, and $\mathbf{s}_u$ is the corresponding patch of the noise-free image. The processing flowchart of sparse coding methods can be summarized into three steps: 1) for each pixel, collect the square window patch and arrange it in a vector form; 2) all the patches are used for sparse coding to obtain the filtered results with perfect sparse signal estimation; 3) aggregate the filtered patches to the original locations with data fusion. Finally, the filtered phase image can be obtained in this manner.

*Experimental analysis:* Here, two sparse regularization algorithms are employed to illustrate the performance of sparse methods. First, the experiments based on aforementioned RADARSAT-2 data are performed using the sparse TV [23] and wavelet [100] regularizations. Fig.13 shows the filtered results that both the algorithms can effectively reduce the noise in the urban areas. Next, the wavelet regularization is applied on the aforementioned interferometric data of Mount Etna and the filtered results are shown as Fig.14.

For clarity, Fig.15 shows more details of the results of the aforementioned algorithms. As Fig.15 shows, Lee filter introduces some artifacts of fringes ambiguous and broken in the areas of high topography and low coherence. In comparison of wavelet filter, region growing and sparse wavelet methods can improve the performance of interferometric phase noise reduction in smooth areas and effectively suppress the phase residues of grainy noise. The sparse wavelet method almost does the best performance in dealing with blurring and discontinuities of phase fringes. More details can be referred to [100].

*Comments:* In addition to the sparse methods, some other advanced technologies have also been tried in InSAR phase denoising, such as random Markov Random Field (MRF) [105-108], tensor decomposition [109] and convolutional neural networks (CNN) [110] and optimization integration with phase unwrapping [111-115]. Roughly, most of them can be classified as the machine learning or deep learning methods and the details are not introduced in this paper.

## VII. CONCLUSION AND PROSPECTS

In this paper, we have reviewed the InSAR phase denoising technology. In the past few decades, a large number of methods have been proposed with great achievements, which are classified into four categories in this paper: 1) traditional local filters; 2) transformed-domain filters; 3) NL filters; and (4) newly advanced methods. More importantly, we have introduced the concepts and surveyed the strengths and limitations of these methods, both theoretically and experimentally. At present, the research achievements on InSAR phase denoising look exciting, which have been successfully applied in the current radar sensors. However, there is still improvement space for the current research in some fields, such as the incoming technology of NL filters and upcoming technology of newly advanced methods. To the best of our knowledge, we try to provide some suggestions on the research tendency, indicating the potential directions in the future.

### A. High efficiency and high precision algorithms

The group of NL filters uses the patch-wise based strategy for sample selection and iterative manner to improve the estimation performance. The filter parameters, such as search window size, patch size, similarity measure criteria and iteration number, are crucial to the efficiency and precision of algorithm implementation in practice. In particular, the complicate texture of interferometric images, known as high topography variation and high heterogeneity, is a high challenge in the face of the NL filters, hindering their practical application [116]. Therefore, how to improve the performances of robustness and efficiency is a very important issue for NL filters, even for all the phase denoising methods.

### B. Sparse methods

The performance of sparse methods is highly dependent on the model of a sparse prior. The sparse representation of interferometric images is still an open question. It is hard to evaluate the sparse solution of interferometric phase: 1) the sparse results depend on the sparse signal recovery technique, such as sparse regularization and sparse coding; 2) it is difficult to distinguish the interferometric phase from noise in the sparse space because there is no explicit theoretical support. Therefore, how to design a well-established paradigm of sparse phase denoiser is significantly important to promote the practical development of this class technology. Meanwhile, high efficiency is also an important factor.

### C. Evaluating phase denoising performance

It is commonly accepted that it is still an open question to objectively evaluate the phase denoising performance. The purpose of phase denoising is to reduce the noise and preserve the phase details as much as possible. In the flat area, noise reduction is the most important issue. In the area of high topography variation, the preservation of phase details without introducing artifacts becomes more important. In addition, the computational complexity is also an important issue. Therefore, it needs further study on how best to comprehensively evaluate the phase denoising performance, which is very helpful to the architecture development of InSAR signal processing.

Phase denoising technology plays an important role to be a mandatory step in InSAR signal processing. Its ultimate objective is to satisfy the practical demands along with the development of new InSAR systems. The purpose of this paper is to provide necessary guideline to the researchers with some novel inspirations. It is our hope that this paper is helpful to promote the further development of InSAR technology.

## ACKNOWLEDGMENTS

We would like to thank the editors and anonymous reviewers for their valuable comments, which have helped to improve the quality of this paper. This work was supported in part by the National Natural Science Foundation of China (NSFC) under Grant 61701106 and Grant 61861136002, in part by the Natural Science Foundation of Jiangsu Province under Grant BK20170698, in part by Shanghai Aerospace Science and Technology Innovation Fund under Grant SAST2018-044. The RADARSAT-2 data were provided by MacDonald, Dettwiler and Associates Ltd online in a public share.

## AUTHORS

**Gang Xu** (M'16) received the B.S. degree and the Ph.D degree in electrical engineering from Xidian University, Xi'an China, in 2009 and 2015, respectively. From 2015 to 2016, he was a full-time Postdoctoral Research Fellow with the School of Electrical and Electronic Engineering (EEE), Nanyang Technological University (NTU), Singapore. He is currently an Associate Professor with the State Key Laboratory of Millimeter Waves, School of Information Science and Engineering, Southeast University, Nanjing, China. His major research interests include high-resolution radar imaging and SAR/ISAR interferometric processing.

**Yandong Gao** received the B.S. and M.S. degrees in survey and mapping engineering from University of Science and Technology Liaoning, Anshan, China, in 2013 and 2016, respectively. He received the Ph.D. degree in geodesy and surveying engineering from China University of Mining and Technology, Xuzhou, China, in 2019. He is currently a Postdoctoral Research Fellow with the School of Environment Science and Spatial Informatics, China University of Mining and Technology, Xuzhou, China. His major research interests include the interferometric phase filtering, phase unwrapping, and synthetic aperture radar interferometry signal processing.

**Jinwei Li** received the M.S. degree and the Ph.D degree in electrical engineering from Xidian University, Xi'an China, in 2009 and 2015, respectively. From 2016, he working as a member at Xi'an institute of space radio technology, Xi'an, China. His major research interests include the fields of interferometric phase filtering, phase unwrapping, and synthetic aperture radar interferometry signal processing.

**Mengdao Xing** (M'04) was born in Zhejiang, China, in November, 1975. He received the Bachelor degree and the Ph.D degree in electrical engineering from Xidian University, Xi'an, China, in 1997 and 2002, respectively. He is currently a full professor with the National Laboratory of Radar Signal Processing, Xidian University, Xi'an, China. His major research interests include SAR, ISAR and over the horizon radar (OTHR).